\documentclass{kapproc} 

\usepackage{procps} 

\usepackage[dvips]{graphicx}

\upperandlowercase

\kluwerbib 

\begin{document}

\articletitle[Bar-driven fueling of galactic nuclei: a 2D view]
{Bar-driven fueling of galactic nuclei: a 2D view}

\author{Eric Emsellem\altaffilmark{1}} 
 
\affil{\altaffilmark{1}Centre de Recherche Astronomique de Lyon,
9 av. Charles Andr\'{e}, 69561 Saint Genis Laval, France}

\begin{abstract}
I briefly discuss evidences for bar-driven gas fueling
in the central regions of galaxies, focusing on scales down to about 10~pc.
I thus mention the building of inner disks, and the link with resonances, as
well as the corresponding kinematic signatures such as $\sigma$-drops and
counter-rotating nuclear disks as probed via integral-field spectroscopy.
\end{abstract}


\section{Introduction}

Before embarking onto this short report on bar-driven fueling of the central
regions of galaxies, let me define what I mean 
by "fueling galactic nuclei". Going from the kpc scale down to the
presumed black hole of a galaxy, we progress through regions where the physical
scales and regimes of the involved processes vary considerably. We should therefore 
not expect a direct link between the large-scale dynamical processes (e.g. the presence
of a bar), and the central engine (accretion disk surrounding the black hole). I would
also like to follow the nomenclature advocated by Jean-Luc Nieto, and thus only
use the words "nucleus" and "nuclear" for structures at 
a scale of the order of $\sim 10$~pc (an excellent illustration 
being the nearby nucleus of M~31). 
I will therefore focus here on the question of "how to accumulate mass at the
scales of galactic nuclei, i.e. in the central tens of parsecs" (see also
Emsellem 2004).

We know that bars can be efficient at redistributing gas within the stellar disk
of a galaxy: outwards when the gas is
in between the outer Lindblad resonance and Corotation, and inwards when it is
inside Corotation (and outside the inner Lindblad resonance - ILR - if there is one).
We therefore expect structures to build up at a scale related to the presumed ILR.
The questions are therefore: do we observe such features, and are they small
enough to be considered as "nuclear" structures?

\section{Building inner disks}

\begin{figure}[ht]
\centerline{\includegraphics[width=\textwidth]{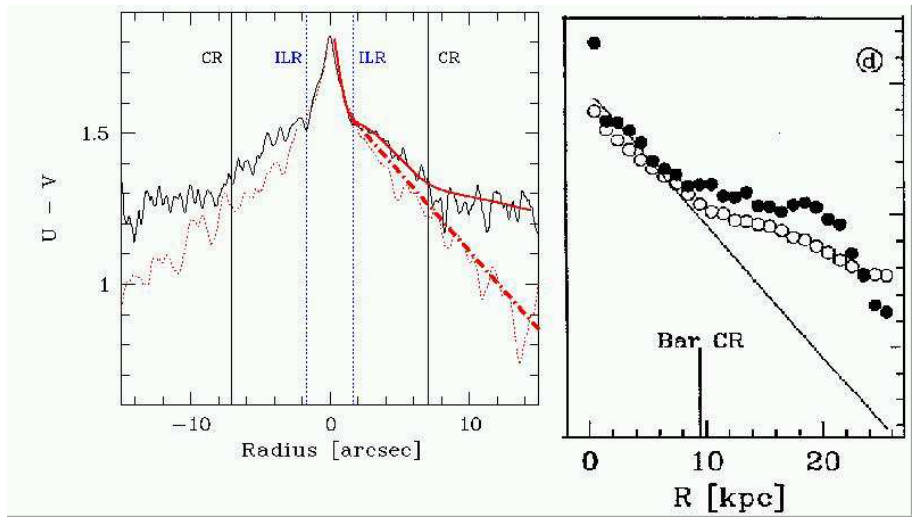}}
\caption{Left panel: mean abundance gradients obtained after redistribution due
to the presence of a bar (from Friedli, Benz \& Kennicutt 1994). The slopes are
clearly different inside and outside the bar. Right panel: $V-I$ colour gradients
along the minor and major axis of NGC~4570. The observed gradients are linked
with the presumed locations of the resonances as expected.}
\label{fig:n4570_1}
\end{figure}

\begin{figure}[ht]
\centerline{\includegraphics[width=\textwidth]{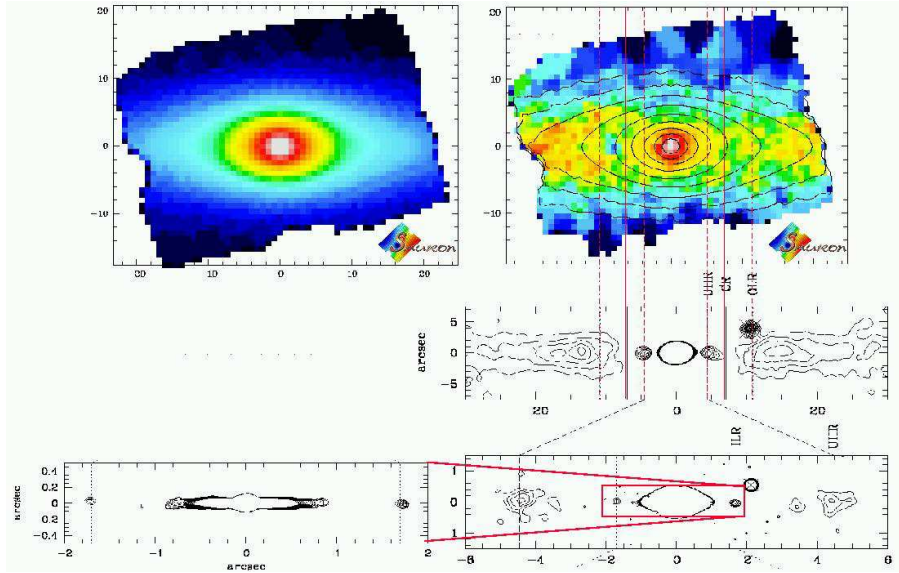}}
\caption{Top panels: SAURON reconstructed intensity (left) and Mgb (right)
maps (to appear in a forthcoming paper of the SAURON team). 
The structures observed in the Mgb map are clearly correlated with the 
presumed resonances, emphasized in the isophotes of an unsharp masking image
of NGC~4570 (contours plots extracted from van den Bosch \& Emsellem 1998).}
\label{fig:n4570c}
\end{figure}

There are numerous examples of galaxy structures which have obviously been formed under
the influence of a bar, the most generic ones being galactic rings (see e.g. work by Buta
and collaborators). I would however like to focus on another type of dynamically cold
systems, namely the building of inner disks. Although the inner disk of e.g., the 
Sombrero galaxy qualifies as a secularly evolved structure (Emsellem 1995), one of the best
case to date remains the photometric features exhibited by the nearly
edge-on S0 galaxy NGC~4570: there is strong evidence that the two ring-like structures and
the 100~pc inner disk of that galaxy are the result of bar-driven secular
evolution (van den Bosch \& Emsellem 1998). Numerical simulations by Friedli, Benz \& Kennicutt
(1994) predicted that a radially decreasing initial abundance gradient should
evolve due to the presence of a bar, with a strong flattening of the gradient
outside cororation, a weakened one inside the bar region and a 
possible starburst at the very centre. This seems consistent with the
observed colour gradients in NGC~4570 along the minor-axis, which retains the
original vertical gradient, and along the major-axis which shows a clear
correlation with the presumed location of the bars and its resonances (Fig.~\ref{fig:n4570_1}).
The jump from gas to stellar abundances, and the use of broad
band colours to assess these gradients should forbid us to conclude too
hastily. However, we recently
obtained line-strength maps of NGC~4570 within the course of the SAURON survey
(Bacon et al. 2001, de Zeeuw et al. 2002, Emsellem et al. 2004), in which we see a
direct correspondence between the previously observed structures and the metal
enrichment (as probed here by e.g. the Mgb index, Fig.~\ref{fig:n4570c}).

\section{Towards the nucleus}

Inner disks with sizes ranging from 100 to 500~pc, such as the one observed in NGC~4570,
are quite common in early-type galaxies and could indeed be the result of
bar-driven accretion followed by star formation (e.g. see the case of NGC~3115).
This would in fact require inner bars with diameter from $\sim 200$ to 1000~pc,
similar to the ones now routinely observed in disk galaxies 
(Laine et al. 2002, Erwin \& Sparke 2002). In some cases, such as the S0/a
galaxy NGC~2974, the size of the fueled region (ILR) is less than 20~pc in
diameter, a scale at which we can start making the link with the nucleus itself
(Emsellem, Goudfrooij \& Ferruit 2003).

\begin{figure}[ht]
\centerline{\includegraphics[width=\textwidth]{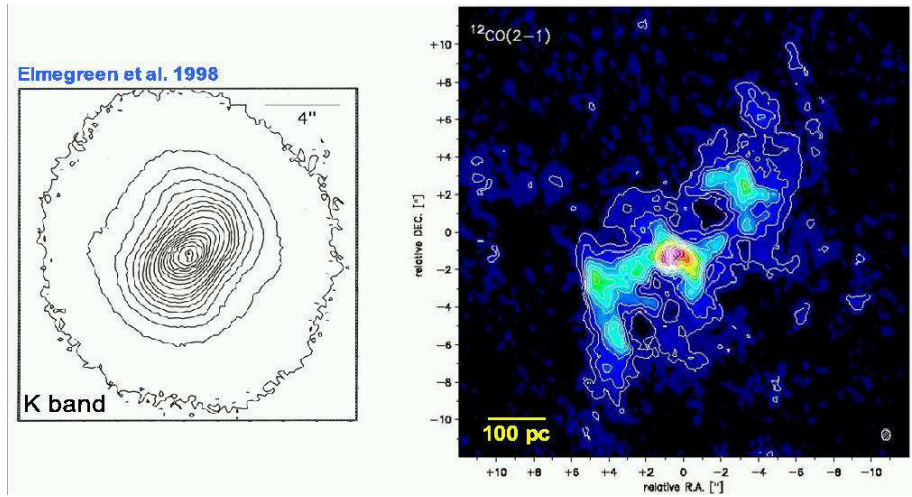}}
\caption{Right panel: $^{12}$C0(2-1) map obtained with tha IRAM interferometer (from Schinnerer 
et al., in preparation) of the central kpc of NGC~6946, showing a two-arm
spiral-like structure reminiscent of dust lanes in bars. Left panel: K band
image (from Elmegreen et al. 1998) showing the presence of an inner bar in
NGC~6946.}
\label{fig:n6946}
\end{figure}

Another clear case of gas fueling of the nuclear regions is provided by
the high resolution $^{12}$CO(2-1) map of NGC~6946 (Schinnerer et al., in preparation;
Fig.\ref{fig:n6946}). The spiral-like distribution of the molecular gas is
reminiscent of the dust lanes observed in barred galaxies, and indeed a small
inner bar has been detected via K band imaging in this galaxy (Elmegreen et al.
1998). The amount of gas which is fueled within the central 20~pc is uncertain, but
star formation is already ongoing there, which may thus lead to the formation of
a flattened nuclear disk. 

\section{Signatures and the importance of bars}

Looking for signatures of past accretion events, we should turn away from purely
photometric features. The dynamical status of the central regions of galaxies
may help us to probe such accretion events long after the nucleus
itself has been formed. The so-called $\sigma$-drops
(the DEBCA project: Emsellem et al. 2001), a central depression in the stellar 
velocity dispersion profile, are now routinely detected
in disk galaxies (e.g. Marquez et al. 2004) and could
indeed be the signatures we are looking for (Wozniak et al. 2003). 
It would be interesting to examine if there is any link between
the blue nuclei observed in most spiral galaxies (e.g. B\"oker et al.
2004) and the presence of such $\sigma$-drops. \\

It should finally be made clear that bars are not the only way gas can
be transported inside the central 10~pc or so. 
The first ingredient for a successful fueling is obviously the availability of gas.
In this context, interactions and/or external accretion have certainly a significant
role in galaxy evolution (either with companions, or from large-scale
structures; see Bournaud's, and Combes' contributions, these Proceedings). This 
is now clearly witnessed in NGC~7332, for which two-dimensional SAURON spectrography
has been obtained (Falc\`on-Barroso et al. 2004). The ionized gas distribution and
kinematics does not leave any doubt on the external origin of part of 
the dissipative component which is counter-rotating with respect to the stars. 
Evidences for the presence of a strong bar are also very strong: 
e.g. a boxy bulge, a cylindrical stellar velocity field. An hybrid scenario
emerges in this case where the interaction with a companion provides the gas
which is then fueled to the central inner 50~pc with the help of a bar (the
formation of which could have been triggered by the interaction). It is
finally important to note that the central stellar kinematics shows
the presence of a counter-rotating stellar disk, less than 100~pc in diameter,
a possible remnant of a past accretion episode.

I would like here to thank my collaborators, and more specifically 
Torsten B\"oker, Eva Schinnerer, Ute Lisenfeld, as well as
the DEBCA and SAURON teams.

\end{document}